\documentclass[
	aps,
    pra,
    superscriptaddress,
    groupedaddress,
    twocolumn,
    a4paper,
    floatfix,
    raggedbottom,
    longbibliography
]{revtex4-2}

\usepackage[american]{babel}
\usepackage[utf8x]{inputenc}

\usepackage{grffile} 

\usepackage{newtxtext,newtxmath}
\usepackage{microtype}

\usepackage{amsmath}
\usepackage{bm}
\usepackage{mathtools}
\usepackage{braket}

\usepackage{graphicx}
\usepackage[svgnames]{xcolor}

\usepackage{siunitx} 

\usepackage{ragged2e}
\usepackage{array}
\usepackage{tabularx}
\usepackage{booktabs}

\definecolor{cset-aps-blueberry}{RGB}{28,128,158}
\definecolor{cset-aps-blue}{RGB}{46,44,184}
\definecolor{cset-aps-turquoise}{RGB}{0,67,88}
\definecolor{cset-aps-limegreen}{RGB}{190,219,67}
\definecolor{cset-aps-green}{RGB}{31,138,112}
\definecolor{cset-aps-yellow}{RGB}{255,225,25}
\definecolor{cset-aps-orange}{RGB}{253,116,0}
\definecolor{cset-aps-red}{RGB}{219,0,43}

\usepackage{tikz}

\usepackage{pgfplots}
\pgfplotsset{%
    every axis legend/.append style={%
        cells={anchor=west},
        at={(0.96,0.04)},
        anchor=south east,
        font=\scriptsize,
        },
    every axis/.append style={%
        yticklabel style={%
            /pgf/number format/fixed zerofill,
            /pgf/number format/precision=2},
        },
    width= \textwidth,
    height=8cm,
    xmajorgrids=true,
    xminorgrids=false,
    minor x tick num=1,
}
\mathtoolsset{showonlyrefs=true}

\newcommand{\Secref}[1]{Sec.~\ref{#1}}
\newcommand{\Eq}[1]{Eq.~\eqref{#1}}
 
\newcommand{\Fig}[1]{Fig.~\ref{#1}} 
\newcommand{\App}[1]{Appendix~\ref{#1}}

\usepackage{hyperref}
\hypersetup{%
    colorlinks=true,
    linkcolor={cset-aps-red},
    linkbordercolor={cset-aps-red},
    filecolor={cset-aps-orange},
    filebordercolor={cset-aps-orange},
    citecolor={cset-aps-blue},
    citebordercolor={cset-aps-blue},
    urlcolor={cset-aps-green},
    urlbordercolor={cset-aps-green},
    menucolor={cset-aps-limegreen},
    menubordercolor={cset-aps-limegreen},
    breaklinks=true,
    pdfborderstyle={/S/U/W 2},
    pdfpagemode=UseOutlines,
    pdfstartpage={1},
}

\begin{document}

\title[Title]{Generalized gravity-gradient mitigation scheme}
\collaboration{Published as
    \href{https://link.aps.org/doi/10.1103/PhysRevA.103.023305}
    {Physical Review A \textbf{103}, 023305 [2021]}}

\author{Christian Ufrecht}
\email{christian.ufrecht@gmx.de}

\address{\vspace{0.2cm}Institut f{\"u}r Quantenphysik and Center for Integrated Quantum Science and Technology (IQ\textsuperscript{ST}), Universit{\"a}t Ulm, Albert-Einstein-Allee 11, D-89069 Ulm, Germany}

\begin{abstract}
A major challenge in high-precision light-pulse atom interferometric experiments such as in tests of the weak equivalence principle is the uncontrollable dependency of the phase on initial velocity and position of the atoms in the presence of inhomogeneous gravitational fields. To overcome this limitation, mitigation strategies have been proposed, however, valid only for harmonic potentials or only for small branch separations in more general situations.
Here, we provide a mitigation formula for anharmonic perturbation potentials including local gravitational effects that vary on length scales much smaller than the spatial extent probed by the atoms and originate e.g.~from buildings that surround the experiment.
Furthermore, our results are applicable to general interferometer geometries with arbitrary branch separation and allow for compensation of Coriolis effects in rotating reference frames.
\end{abstract}

\maketitle

\section{Introduction}
The high sensitivity of  light-pulse atom interferometry with promising applications as inertial sensor in gravimetry \cite{DroppingAtoms,Syrte}, gradiometry \cite{Gradiometrie1,Gradiometrie3,Gradiometrie4} and tests of the weak equivalence principle (WEP) \cite{WEP1,WEP3, WEP6} has led to ambitious proposals on  ground \cite{WEPProposalVLBAI,WEPProposalGround, RedshiftUfrecht, RouraRedshift} and in space \cite{WEPSTEQUEST,WEPSatelliteProposal}.

A serious challenge for next-generation atom interferometric high-precision measurements is posed by non-linearities in the gravitational potential and Coriolis forces
which lead to non-perfect overlap of the trajectories in both momentum and position after the final laser pulse. As a consequence, the phase contains the initial position and velocity of the atoms and the contrast drops dramatically in long-time interferometry \cite{WEPSTEQUEST,Compensation3}. Since control over the initial conditions is limited \cite{Nobili}, mitigation strategies had to be developed.
Interferometer schemes insensitive to initial kinematics in the presence of homogeneous gradients and rotations can be constructed by folding the interferometer geometry symmetrically \cite{Marzlin1, Marzlin2}. However, in these schemes also the dominant part of the phase from linear gravity cancels, including a possible violation signal in a test of the WEP.

Similar to the mitigation strategies developed for Coriolis effects by using tip-tilt mirrors \cite{Peters,Hogan,CompensationRotation1}, the crucial insight to achieve compensation of initial-condition-dependent phases in the presence of homogeneous gravity gradients  was a
modification of the pulse timing \cite{Compensation3} or the momentum transfer of the central pulse e.g.~in a Mach-Zehnder (MZ) interferometer as a function of the gradients \cite{Compensation1,Compensation4,Compensation2}. This method reduces the mismatch of the trajectories at the end of the interferometer sequence while leaving the phase from linear gravity unaffected.
Besides other successive work to mimic an inertial frame \cite{Compensation7,Compensation6}, gravity-gradient compensation was extended to spatially inhomogeneous gravity gradients \cite{Compensation2} based on an additional modification of the momentum transfer of the final laser pulse.
However, the derivation in Ref.~\cite{Compensation2} is based on the midpoint theorem \cite{Borde3}, which becomes less accurate with increasing branch separation when applied to anharmonic potentials. Therefore, future proposals including large-momentum-transfer techniques \cite{FineStructure4,Bloch1} to increase the space-time area of the interferometer or long-time interferometry with large branch separation require a further generalization of the method to such situations.

In this article particular emphasis is put on the perturbative character of the description which allows the application of the mitigation scheme to arbitrary anharmonic perturbations in the gravitational potential as for example present in the experimental setups of Refs.~\cite{Compensation2, VerticalProfile}.
The formula derived in the present article is furthermore valid for general interferometer geometries and arbitrary branch separation.
Our derivation within a full quantum-mechanical framework also allows to consistently include contributions from wave-packet dynamics.

In \Secref{Path-dependent description} we briefly review the perturbative formalism 
employed to derive the compensation formula for inhomogeneous gravity gradients in \Secref{General cancellation scheme}. In \Secref{Rotations} we generalize the formula to include rotations and finally discuss conditions for validity of our derivation in \Secref{Discussion and validity}.

\begin{figure}
	\begin{center}
		\includegraphics[width=\linewidth]{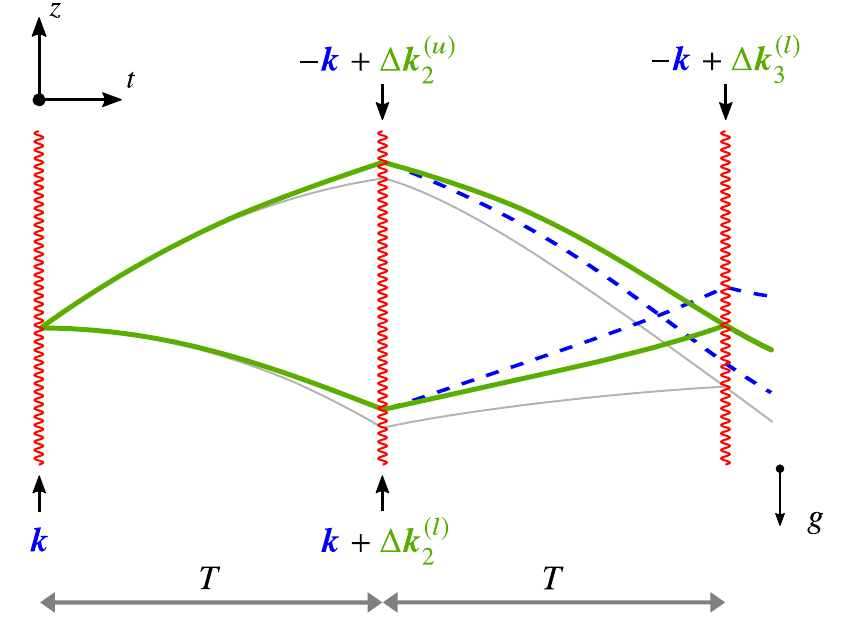}
		\caption{\textit{Gravity-gradient compensation in a small cubic potential.} In an MZ interferometer a $\pi/2$ pulse at $t=0$ splits the initial wave function into two components and transfers the momentum $\hbar k$ on one of them. The atoms are then redirected by a $\pi$ pulse at $t=T$ and finally the wave function is recombined by a second $\pi/2$ pulse at $t=2T$. In absence of a non-linear gravitational potential the atoms follow the unperturbed trajectories (thin solid lines). These trajectories are modified (dashed lines) in presence of a small perturbation potential so that the branches do not overlap perfectly in both momentum and position at the final laser pulse, the interferometer is then referred to as open. Note that this deviation is displayed strongly exaggerated in the figure. If the wave vector of the second and the final pulse is modified appropriately by $\Delta \bm{k}_\ell$ (decreased in the example shown in the figure), the interferometer can be closed (thick solid lines) and dependence of the phase on the initial conditions is eliminated to first order.
		}
		\label{fig:MZ}
	\end{center}
\end{figure}

\section{Path-dependent description}
\label{Path-dependent description}
In this article, we rely on the perturbative formalism recently developed in Refs.~\cite{Ufrecht, PAI}. The Hamiltonian
\begin{equation}
\label{Hamiltonian}
\hat{H}^{(\alpha)}=\hat{H}_0^{(\alpha)}+V(\hat{\bm{r}},t)
\end{equation}
describes the evolution through the interferometer along the upper ($\alpha=u$) and the lower ($\alpha=l$) branch. It consists of a dominant part $\hat{H}_0^{(\alpha)}$ with respect to which the interferometer is closed (that is perfect overlap after the final pulse) and a perturbation $V(\hat{\bm{r}},t)$ which slightly opens the interferometer and renders the phase dependent on initial position and velocity of the atoms. 
As an example we illustrate in \Fig{fig:MZ} the case of an MZ interferometer where a small cubic potential leads to a slight mismatch of the trajectories at the final laser pulse and explain how custom-tailored laser pulses can mitigate this effect.

In a gravimeter configuration the natural choice for the unperturbed Hamiltonian is
\begin{equation}
\label{UnperturbedHamiltonian}
\hat{H}_0^{(\alpha)}=\frac{\hat{\bm{p}}^2}{2m}-m\bm{g}\hat{\bm{r}}+ V_\mathrm{em}^{(\alpha)}( \hat{\bm{r}},t)
\end{equation}
where $m$ is the mass of the atoms and $\bm{g}$ is the local linear acceleration. Furthermore,  
 \begin{equation}
 \label{LaserInteraction}
 V_\mathrm{em}^{(\alpha)}( \hat{\bm{r}},t)=-\hbar \sum_\ell[\bm{k}^{(\alpha)}_\ell \hat{\bm{r}}+\varphi_\ell^{(\alpha)}]\delta(t-t_\ell)
 \end{equation}
is the effective laser interaction potential, imprinting the momentum $\hbar \bm{k}_\ell$ and the laser phase $\varphi_\ell^{(\alpha)}$ at
time $t_\ell$ on branch $\alpha$. Additional contributions to the gravitational potential such as gradients of Earth's gravitational potential, gravitational fields of the local environment such as from buildings surrounding the experiment \cite{VerticalProfile} etc., are incorporated into the perturbation potential $V(\hat{\bm{r}},t)$ and treated perturbatively.

The phase $\varphi$ and contrast $C$ of a matter-wave interferometer is defined as
\begin{equation}
\label{OverlapOperator}
\langle\hat{U}^{(l)\dagger}\hat{U}^{(u)}\rangle=\langle \mathrm{e}^{\mathrm{i}\hat{\phi}} \rangle= C\mathrm{e}^{\mathrm{i}\varphi}
\end{equation}
where $\hat{U}^{(\alpha)}$ is the time-evolution operator with respect to Hamiltonian \eqref{Hamiltonian} for the respective branch and the expectation value is taken with respect to the initial  wave function. 
In Refs.~\cite{PAI, Ufrecht} we merged the two  time-evolution operators on the left-hand side in favour of the operator $\hat{\phi}$, which reads to first order in the perturbation 
\begin{equation}
\label{phasecalculation}
    \hat{\phi}=\phi_0-\frac{1}{\hbar}\oint\!\mathrm{d}t\,V(\hat{\bm{r}}(t)) 
\end{equation}
and where $\phi_0$ is the interferometer phase for vanishing perturbation.
The integral runs along the upper branch and returns along the lower. It is taken over the perturbation potential evaluated at the branch-dependent and operator-valued Heisenberg trajectories $\hat{\bm{r}}^{(\alpha)}(t)$ generated by the \textit{unperturbed} Hamiltonian
\eqref{UnperturbedHamiltonian}.
 
The Heisenberg trajectories can be decomposed into the sum of the classical trajectory
with the initial conditions given by the initial mean position and velocity of the wave packet and a fluctuation operator of the form   
$ \hat{\overline{\bm{r}}}(t)=\hat{\bm{r}}-\langle\hat{\bm{r}}\rangle+[\hat{\bm{p}}-\langle\hat{\bm{p}}\rangle]t/m$ where the expectation values are taken with respect to the initial wave function \cite{PAI}.  This decomposition is valid as the unperturbed Hamiltonian is only linear in the position operator. 
The standard deviation  of the fluctuation operator is a measure for the size of the expanding wave packet and characterizes wave-packet effects.
In a modification to the form from Ref.~\cite{PAI} we define the fluctuation operator
 \begin{equation}
\label{Fluctuation1}
\delta \hat{\overline{\bm{r}}}(t)=\delta \bm{r}_0(t)+\hat{\bm{r}}-\langle\hat{\bm{r}}\rangle+\frac{\hat{\bm{p}}-\langle\hat{\bm{p}}\rangle}{m}t
\end{equation}
 which includes the deviation of the trajectories $\delta \bm{r}_0(t)$ due to uncertainties in the initial conditions. Thus, the Heisenberg trajectory reads
  \begin{equation}
 \label{Heisenberg}
     \hat{\bm{r}}^{(\alpha)}(t)=\bm{r}^{(\alpha)}_0(t)+\delta \hat{\overline{\bm{r}}}(t)
 \end{equation}
where $\bm{r}^{(\alpha)}_0(t)$ is the classical unperturbed trajectory without this deviation. 
Consequently, we find for the expectation value that $\langle \delta\hat{\overline{\bm{r}}}(t)\rangle=\delta \bm{r}_0(t)$.
Note that all expectation values are taken with respect to the initial wave function.
In summary, in the real classical unperturbed trajectories $\bm{r}_0^{(\alpha)}(t)+\delta \bm{r}_0(t)$ we choose $\bm{r}_0^{(\alpha)}(t)$ as a fixed reference while $\delta \bm{r}_0(t)$ describes their uncertainty due to e.g.~limited initial characterization time \cite{Nobili}.

Making use of the decomposition from \Eq{Heisenberg}, a small value of $\delta\bm{r}_0(t)$ and a small wave-packet size will allow for a Taylor expansion of the perturbation potential around the classical unperturbed trajectories in the next section.
The dominant linear contribution of  $\delta \hat{\overline{\bm{r}}}(t)$ in $\hat{\phi}$ not only introduces a dependence on the initial conditions but also leads to a loss of contrast \cite{WEPSTEQUEST,Compensation3} when calculating the expectation value in \Eq{OverlapOperator}.
Following Refs.~\cite{Compensation1}, slightly modifying the momentum transfer of the laser pulses will
eliminate $\delta \hat{\overline{\bm{r}}}(t)$ to leading order and therefore strongly mitigate these two effects.

In our fully quantum-mechanical treatment phase contribution that arise from the square and higher powers of $\delta \hat{\overline{\bm{r}}}(t)$ include both the residual, strongly suppressed dependence on the initial conditions and wave-packet effects due to e.g.~different dynamics along the interferometer branches.

\section{Gravity Gradients}
\label{General cancellation scheme}
As anticipated in the previous section, we start by replacing
 \begin{equation}
 \label{changek}
     \bm{k}_\ell^{(\alpha)}\rightarrow \bm{k}_\ell^{(\alpha)}+\Delta\bm{k}_\ell^{(\alpha)}
 \end{equation}
for each laser pulse and modify the effective laser-atom interaction potential in \Eq{LaserInteraction} accordingly. As the validity of the perturbative approach requires a closed unperturbed interferometer \cite{PAI}, we keep the unperturbed Hamiltonian \eqref{UnperturbedHamiltonian} unchanged and therefore consider 
 \begin{equation}
\label{Perturbationk}
   \Delta V _\mathrm{em}^{(\alpha)}(\bm{\hat{r}},t)= -\hbar \sum_\ell\Delta\bm{k}^{(\alpha)}_\ell \hat{\bm{r}}\delta(t-t_\ell)
\end{equation}
as part of the perturbation potential $V(\hat{\bm{r}},t)$.
In the following we will omit the branch index $\alpha$ whenever possible.
Furthermore, quantities without operator hat are understood to be evaluated at the unperturbed trajectories $\bm{r}_0(t)$ and we will omit the explicit time dependence.

Inserting \Eq{Heisenberg} into the perturbation potential followed by Taylor expansion about the classical trajectories, we find $V(\hat{\bm{r}})=V-m \bm{a}\delta\hat{\overline{\bm{r}}}+...$  with the acceleration $\bm{a}=-\nabla V/m$ and consequently 
 to first order in $\delta\hat{\overline{\bm{r}}}$
\begin{equation}
\label{DerivationCancellation}
 \hat{\phi} =\phi_0 -\frac{1}{\hbar}\oint \!\mathrm{d}t\,\left(V -m\bm{a}\delta\hat{\overline{\bm{r}}}\right)+\Delta \phi_k+\sum_\ell \Delta \bm{k}_\ell \delta\hat{\overline{\bm{r}}}(t_\ell)\,.
\end{equation}
The last contribution in \Eq{DerivationCancellation} and the phase
\begin{equation} 
\Delta \phi_k=\sum_\ell\Delta\bm{k}_\ell^{(u)}\bm{r}_0^{(u)}(t_\ell)-\Delta \bm{k}_\ell^{(l)}\bm{r}_0^{(l)}(t_\ell)
\end{equation}
 originate from the perturbation $\Delta V_\mathrm{em}$ evaluated at Heisenberg trajectories and  we abbreviated $\Delta \bm{k}_\ell=\Delta \bm{k}^{(u)}_\ell-\Delta \bm{k}^{(l)}_\ell$  to alleviate notation.

Thanks to the simple form of the unperturbed Hamiltonian we find $\delta \bm{r}_0=\delta\bm{r}_\mathrm{i}+\delta\bm{v}_\mathrm{i} t$ where $\delta\bm{r}_\mathrm{i}$ and $\delta\bm{v}_\mathrm{i}$ are the uncertainties of initial position and velocity. Thus, making additionally use of the explicit form of the fluctuation operator shown in \Eq{Fluctuation1},
 all terms in \Eq{DerivationCancellation} linear in $\delta \hat{\overline{\bm{r}}}$  vanish if we require that
\begin{equation}
\label{LinearEquation}
    \bm{J}_0=-\sum_\ell\Delta \bm{k}_\ell \;\;\;\text{and}\;\;\; \bm{J}_1=-\sum_\ell\Delta \bm{k}_\ell t_\ell
\end{equation}
with the abbreviations
\begin{equation}
\label{DefinitionJ}
    \bm{J}_0=\frac{m}{\hbar}\oint\!\mathrm{d}t\, \bm{a}(t) \;\;\;\text{and}\;\;\; \bm{J}_1=\frac{m}{\hbar}\oint\!\mathrm{d}t\,\bm{a}(t)t \,.
\end{equation}
After eliminating the operator-valued terms in \Eq{DerivationCancellation}, the operator $\hat{\phi}$ has become a $c$-number (to the order considered here) and we find
\begin{equation}
\label{dominantPhase}
\varphi=\phi_0 -\frac{1}{\hbar}\oint \!\mathrm{d}t\,V +\Delta \phi_k
\end{equation}
where we stress again that $V$ is evaluated at the classical unperturbed trajectories $\bm{r}_0^{(\alpha)}$.
The linear set of equations in \Eq{LinearEquation} can be solved in general  if we slightly change the wave vectors of the laser at two different times, say $t_1$ and $t_2$, so that we find
\begin{equation}
\label{ks}
\Delta \bm{k}_1=-\frac{\bm{J}_1 - \bm{J}_0 t_2}{t_1-t_2}\;\;\;\;\;\text{and}\;\;\;\;\;\Delta \bm{k}_2=\frac{\bm{J}_1 - \bm{J}_0 t_1}{t_1-t_2}\,.
\end{equation}
The functions $\bm{J}_0$ and $\bm{J}_1$ in \Eq{DefinitionJ} allow an intuitive interpretation. The former is proportional to the integrated differential acceleration between the branches, while the latter corresponds to its average over time.
As shown in \App{Approximations for smooth gradients} and \ref{Special form of gradients}, the dependence of $\Delta \bm{k}_1$ and $\Delta \bm{k}_2$ on these quantities allows to design interferometer geometries for which the mitigation scheme simplifies.
Furthermore, in \Eq{ks} any two distinct laser pulses can be chosen that
not necessarily have to correspond to the second and final laser pulse.

 Once $\Delta \bm{k}_1$ and $\Delta \bm{k}_2$ are known, the shifts in momentum transfer can be distributed between the two branches satisfying $\Delta \bm{k}_\ell=\Delta \bm{k}_\ell^{(u)}-\Delta \bm{k}_\ell^{(l)}$. For example we find in case of a laser pulse imprinting opposite momentum on the two branches that
$\Delta\bm{k}_\ell^{(u)}=-\Delta \bm{k}_\ell^{(l)}=\Delta \bm{k}_\ell/2$ as for instance in case of the central pulse in \Fig{fig:MZ}.

Before we generalize \Eq{ks} to rotating frames in the next section, we conclude by the following remarks.

The key advantage of the mitigation scheme is that the accuracy to which initial momentum and position of the atoms need to be determined is significantly relaxed, which might otherwise take longer than the experiment itself in future satellite-based WEP tests \cite{Nobili}.

After mitigation of the initial kinematics the phase still depends on the local gradients \cite{NobiliPRR} (second and third term on the right-hand side of \Eq{dominantPhase}) and it was discussed whether it is meaningful to extent the mitigation scheme to also compensate these contributions \cite{RouraDubetsky1, RouraDubetsky2}.
While in principle compensation of the initial conditions can be achieved without knowledge of the gravitational background by calibrating the interferometer prior to the measurement \cite{WEP6,Compensation2}, extension of the mitigation scheme to cancel all gradient-dependent phases would still require a precise characterization of the gravitational background as in classical tests with torsion balances \cite{ClassicalCharacter}. It therefore seems more practical to postcorrect these phases.
This postcorrection of course can only be done to the accuracy by which the gravity gradients are known. Therefore, it needs to be guaranteed that the remaining phase shifts only influence the measurement result below the target accuracy.

In atom interferometric tests of the WEP the phases of two interferometers operated with different atomic isotopes or atomic species are compared. In the latter case the wave numbers of the lasers are generally  different. In order to compare the differential effective gravitational acceleration, the phases have to be rescaled by the respective wave numbers (assuming the same interferometer time $T$) before taking the difference.
In ground-based tests this procedure is only meaningful as long as the uncertainty in the wave vectors is smaller than the target accuracy of the WEP violation parameter. In microgravity, however, this constraint is significantly relaxed.
If the mitigation scheme is applied, remaining gradient-induced phase shifts independent of the initial kinematics cancel differentially in case of homogeneous gradients. In case of locally varying gravitational potentials, however, the atoms feel different local potentials along the species-dependent trajectories. As a consequence, these phase contributions are only suppressed in the differential rescaled phase but not cancelled. 
While this remaining differential phase is small, it might nevertheless impose limits on future tests of the WEP on ground if the gravitational background is not known precisely.

In \Eq{DerivationCancellation} the Taylor expansion is truncated at first order in $\delta \hat{\overline{\bm{r}}}$.
Corrections to the phase from higher powers in the fluctuation operator can be calculated with the cumulant expansion \cite{Cumulants2, Ufrecht, PAI}, however, are often negligible \cite{PAI}. Corrections of this kind will be discussed in more detail in \Secref{Discussion and validity}.
As anharmonic potentials are treated quantum mechanically in this work rather than within a semiclassical approximation, our results also cover the application of large-momentum transfer techniques where the branch separation can become comparable to the spatial extent probed by the atoms. 
In \App{Approximations for smooth gradients} we explain the approximations needed to 
obtain the expressions derived in the Supplemental material of Ref.~\cite{Compensation2} and discuss their validity.
We furthermore show how \Eq{ks} reduces to the result originally derived in  Ref.~\cite{Compensation1} for an MZ interferometer in presence of homogeneous  gravity gradients.
In \App{Special form of gradients} we investigate simplifications of our general results in case of trajectories symmetric in time.
We stress the importance of treating the perturbation potential locally \cite{ReplyDubetsky}. For instance the gravitational profile reported in Ref.~\cite{VerticalProfile} cannot be Taylor expanded over the extent of the interferometer due to its  variations on short lengths scales. A numerical integration of \Eq{DefinitionJ} for this example shows that these local perturbations influence the value of $\Delta \bm{k}_1$ at the ten-percent level and above.

\section{Rotations}
\label{Rotations}
For experiments in a rotating reference frame Coriolis and centrifugal forces need to be considered additionally.
Fortunately, it is straightforward to extend our result to such situations as will be shown next.
The Hamiltonian in a rotating frame is obtained by adding
\begin{equation}
    \hat{H}_\Omega = \bm{\Omega}\cdot (\bm{\hat{p}}\times \bm{\hat{r}})
\end{equation}
to Hamiltonian \eqref{Hamiltonian} where $\bm{\Omega}$ is the rotation frequency.
Additional centrifugal forces present for example in a reference frame fixed on Earth's surface only redefine the direction and absolute value of $\bm{g}$.
In complete analogy to \Secref{General cancellation scheme} we integrate $\hat{H}_\Omega$ along the Heisenberg trajectories shown in \Eq{Heisenberg} and recall that $\hat{\bm{p}}(t)=m\,\mathrm{d}/\mathrm{d}t\, \hat{\bm{r}}(t)$. Consequently, \Eq{DerivationCancellation} is extended by the term
\begin{equation}
-\frac{1}{\hbar}\oint\!\mathrm{d}t\,\bm{\Omega}\cdot [\bm{\hat{p}}(t)\times \bm{\hat{r}}(t)]=\phi_\Omega+\frac{2m}{\hbar}\oint\!\mathrm{d}t\,[\bm{v}_0(t)\,\times\, \bm{\Omega} ]\cdot \delta \hat{\overline{\bm{r}}}(t)
\end{equation}
where we neglected terms quadratic in the fluctuation operator and  made use of partial integration for which we appreciated that the unperturbed interferometer is closed. Furthermore, $\bm{v}_0(t)$ is the velocity of the atoms on the unperturbed trajectories and we abbreviated
\begin{equation}
\phi_\Omega=-\frac{1}{\hbar}\oint\!\mathrm{d}t\,\bm{\Omega}\cdot [\bm{p}_0(t)\times \bm{r}_0(t)]\,.
\end{equation}
Consequently, by comparing to \Eq{DerivationCancellation}, the mitigation schemes can be generalized to rotating reference frames with the replacement
\begin{equation}
    \bm{a}(t)\rightarrow \bm{a}(t)+2\bm{v}_0(t)\times \bm{\Omega}
\end{equation} in \Eq{DefinitionJ} and by adding $\phi_ \Omega$ to \Eq{dominantPhase}.
Alternatively, the effects of rotations can be analyzed in a non-rotating frame, where the laser is rotating instead \cite{Kleinert}.

\section{Validity of perturbative treatment}
\label{Discussion and validity}

 In the previous section we developed a general mitigation scheme based on a perturbative treatment, covering both rotations and gravity gradients. In the following we discuss the validity of this approach and the approximations made in the derivation. 

In a perturbative calculation of the phase in powers of the potential $V$ subsequent orders are suppresses by $\epsilon=\Delta V T^2/(m\xi^2)$ \cite{PAI}  where $\Delta V$ is the characteristic change of the potential over the interferometer size, $\xi$ is the typical length scale on which the potential changes and  $T$ the characteristic interferometer time.
The parameter $\epsilon$ can be understood as the deviation of the trajectories caused by the perturbation compared to the length $\xi$.
For gravity gradients on Earth's surface corresponding to the potential $V=m \bm{r}^{\mathrm{T}}\Gamma \bm{r}/2$, one would choose $\xi$ as the extent of the interferometer, given approximately by $\xi=gT^2/2$ in a gravimeter configuration. Thus, $\Delta V\sim m \Gamma \xi^2/2$ and consequently $\epsilon=\Gamma T^2$, leading to a value $\epsilon<10^{-5}$
for typical interferometer times. Local variations as in the gravitational potential of Ref.~\cite{VerticalProfile}, in contrast, can lead to values of $\xi$ much smaller than the spatial extent probed by the atoms.
A similar suppression factor for rotations takes the form $\epsilon_\Omega=\Omega T$ with $\epsilon_\Omega<10^{-4}$ for the rotation of Earth.
Consequently, the relative uncertainty in the phase achieved by a first-order calculation already is of the order of $\epsilon$.

The term
$m \oint \!\mathrm{d}t\, \bm{a}\delta\hat{\overline{\bm{r}}}/\hbar$
in \Eq{DerivationCancellation} introduces the dominant dependence on the initial conditions. Estimating $\nabla V\sim \delta V/\xi$ \cite{PAI} where $\delta V$ is the change of the potential over the branch separation and introducing the abbreviation  $\eta = \delta V T/\hbar$, this phase contribution scales as  $\eta  \delta r_0/\xi$.

Application of the mitigation scheme requires prior knowledge of the gravitational background which can be obtained by measurement, numerical simulation of the gravitational sources surrounding the experiment, or a combination of both. However, determination of deviations from linear gravity will only be possible to some relative uncertainty $\kappa$,  which then also constitutes the suppression factor for initial-condition dependent phases. Estimations suggest that at least $\kappa=10^{-3}$ seems plausible \cite{Compensation2,WEPSatelliteProposal}, thereby considerably relaxing the requirements on determination of initial position and velocity of the atoms. 
As described in the beginning of this section, further terms linear in the initial conditions which would result from the second-order calculation in the perturbation potential are suppressed by $\epsilon$ compared to the first-order terms.
Consequently, an extension of the mitigation scheme to second order in the perturbation \cite{Ufrecht} is only necessary if $\kappa<\epsilon$
since otherwise initial-condition-dependent phases that are compensated only partially are still larger than contributions from the second-order calculation. Higher-order corrections to the dominant phase in \Eq{dominantPhase}, in contrast, can be obtained as shown in detail in Ref.~\cite{PAI}.

Moreover, Taylor expansion of \Eq{phasecalculation} around the classical unperturbed trajectories to first order neglects terms scaling as $\eta (
\delta \hat{\overline{\bm{r}}}/\xi)^2$.  Comparing to the residual contribution $\kappa\,\eta 
\delta \hat{\overline{\bm{r}}}/\xi$ from the first-order calculation, these terms and further corrections can be disregarded if 
$\delta r_\mathrm{i}/\xi< \kappa$ and $\delta v_\mathrm{i} T/\xi< \kappa$.
For $\delta r_\mathrm{i}\sim 1$\textmu m as well as $\delta v_\mathrm{i}\sim 1$\textmu m/s and $\xi\sim 1\,\mathrm{m}$ the latter requirements are satisfied well. 

 In addition, wave-packet effects originating from different expansion dynamics along the branches in an anharmonic potential are generally small but straightforward to include if necessary.

\section{Discussion}
Finally, we conclude by the following remarks.
The acceleration in \Eq{DefinitionJ} not necessarily points in direction of momentum transfer. Consequently, gravity-gradient compensation might also require a tilt of the mirrors in order to adapt the direction of $\bm{k}$ appropriately. However, generally for experiments on Earth's surface
the required modification of momentum transfer orthogonal to the sensitive axis ($\bm{k}$ pointing in direction of $\bm{g}$) often is much smaller than the parallel component due to symmetry in the mass distribution surrounding the apparatus \cite{VerticalProfile}.

Obviously, the compensation method is equally applicable to perturbations of non-gravitational origin.
However, initial condition-dependent phases from e.g.~magnetic field gradients \cite{Magneticshielding}, black-body radiation \cite{BBR1}, etc.~are generally much smaller than those from gravity and can be neglected.

Moreover, to avoid the necessity of a precise characterization of the gravitational background,
the compensation scheme can be implemented experimentally through calibration prior to the measurement by introducing artificial large deviations of the initial conditions \cite{WEP6,Compensation2}.

In the reference frame of an inertial-pointing satellite orbiting Earth the gravitational potential is time dependent.
In a WEP test the varying projection of a possible violation signal on the sensitive axis can be utilized to demodulate systematic effects \cite{WEPSatelliteProposal}. This technique also might relax the required accuracy \cite{Compensation8} to which the gravity gradients have to be measured for application of the mitigation scheme. Note that formula \eqref{ks} also applies to  time-dependent gravitational potentials as in this situation.

\section{Acknowledgements}
The author thanks \'{E}.~Wodey and S.~Loriani for fruitful discussions and W.~P.~Schleich, A.~Roura, A.~Friedrich, F.~Di Pumpo and E.~Giese for a careful reading of the manuscript.
This work is supported by the German Aerospace Center (\href{http://dx.doi.org/10.13039/501100002946}{Deutsches Zentrum für Luft- und Raumfahrt}, DLR) with funds provided by the Federal Ministry for Economic Affairs and Energy (\href{http://dx.doi.org/10.13039/501100006360}{Bundesministerium f\"ur Wirtschaft und Energie}, BMWi) due to an enactment of the German Bundestag under Grant Nos. DLR~50WM1556 and 50WM1956.
The author thanks the Ministry of Science, Research and Art Baden-Württemberg (\href{http://dx.doi.org/10.13039/501100003542}{Ministerium f\"ur Wissenschaft, Forschung und Kunst Baden-Württemberg}) for financially supporting the work of IQ$^\mathrm{ST}$.

\section*{Appendix}

\appendix

\section{Weakly varying potential}
\label{Approximations for smooth gradients}
In this appendix we start from the general result in \Eq{ks} and rederive the result of Ref.~\cite{Compensation2} in case of small branch separation.
The modified wave vector in \Eq{ks} is a function of the atom's mass as the gravitational potential $\Phi$ with $V=m\Phi$ is evaluated at the mass-dependent trajectories. However, if the local acceleration varies only moderately over the branch separation (of the order of centimeter for a few $\hbar k$ momentum transfer and a 10-m baseline) this dependence cancels out and we will find the result of Ref.~\cite{Compensation2} for an MZ interferometer.

\subsection{Local gravity gradients}
 To prove this statement, we first decompose the trajectories 
\begin{equation}
\bm{r}_0^{(\alpha)}(t)=\overline{\bm{r}}(t)+\bm{r}^{(\alpha)}_*(t)
\end{equation}
into a suitably chosen branch-independent mean trajectory $\overline{\bm{r}}(t)$ and the deviation $\bm{r}^{(\alpha)}_*(t)$. Thus, we  Taylor expand
\begin{equation}
\label{TaylorDecomposition}
    \bm{a}^{(\alpha)}(\bm{r}_0)=\bm{a}(\overline{\bm{r}})-\Gamma(\overline{\bm{r}})\bm{r}^{(\alpha)}_*+...
\end{equation}
where the gradient tensor is defined as $\Gamma_{ij}=\partial_i\partial_j V/m$. Substituting \Eq{TaylorDecomposition} into \Eq{DefinitionJ}, we find 
 \begin{figure}
	\begin{center}
		\includegraphics[width=\linewidth]{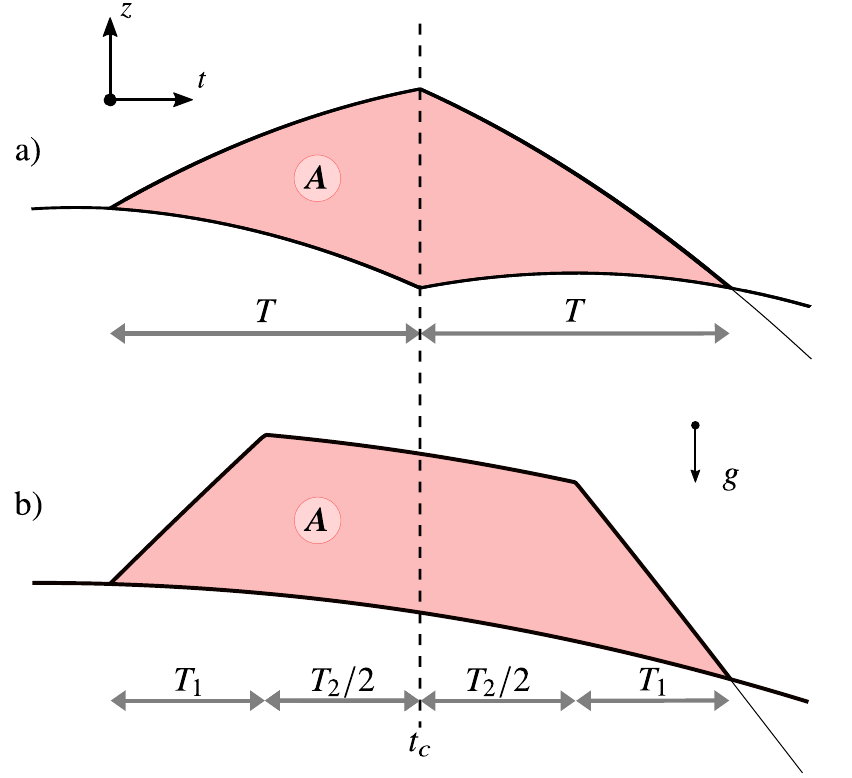}
		\caption{\textit{Compensation for uniform gravity gradients.} 
		a) In an MZ interferometer the position of the geometric center $t_c$ on the time axis of the space-time area $\bm{A}$ enclosed by the two branches coincides with the position of the central pulse. For this reason, as shown in the main text, global gravity gradients can be compensated by adapting the momentum transfer of the central pulse only. The required modification is proportional to the space-time area enclosed by the trajectories.
		b) In a Ramsey-Bord\'{e} interferometer the geometric center is situated exactly in between the two central pulses at $t=T_1$ and $t=T_1+T_2$.
		In this case compensation can be achieved by modifying the momentum transfer of these pulses equally.}
		\label{fig:Centeroid}
	\end{center}
\end{figure}
 \begin{equation}
\label{ApproximateDefinitionJ}
    \bm{J}_0=-\frac{m}{\hbar}\oint\!\mathrm{d}t\, \Gamma(\overline{\bm{r}})\bm{r}_* \;\;\;\text{and}\;\;\; \bm{J}_1=-\frac{m}{\hbar}\oint\!\mathrm{d}t\,\Gamma(\overline{\bm{r}})\bm{r}_*\,t 
\end{equation}
since $\bm{a}(\overline{\bm{r}})$ is independent of the branch and therefore cancels in the looped integrals. 
If the mean trajectory $\overline{\bm{r}}$ only contains the mass-independent part of the  trajectory generated by linear gravity while $\bm{r}_*$ is the additional contribution from the laser pulses, \Eq{ApproximateDefinitionJ} becomes mass independent since $\bm{r}_*$ is inversely proportional to the mass through $\bm{v}_r=\hbar \bm{k}/m$.
To connect with previous results,
we specify the case of an MZ gravimeter where the atoms are launched initially in $z$ direction so that $\overline{\bm{r}}=gt(T-t/2)\bm{\mathrm{e}}_z$.
Consequently, with
\begin{alignat}{3}
&\bm{r}_*^{(u)}=\bm{v}_r t\,,\quad\quad\quad &&\bm{r}_*^{(l)}=0\quad\quad\quad\quad\quad\quad &&0\leq t<T\\  
&\bm{r}_*^{(u)}=\bm{v}_r T\,, && \bm{r}_*^{(l)}=\bm{v}_r (t-T) &&T\leq t\leq2T
\end{alignat}
we find from \Eq{ApproximateDefinitionJ} 
the expressions
\begin{equation}
 \label{Naeherung1}
\bm{J}_0=-\int\limits_0^T\!\mathrm{d}t\,t \Gamma(\overline{\bm{r}})\bm{k} -\int\limits_T^{2T}\!\mathrm{d}t\, (2T-t) \Gamma(\overline{\bm{r}})\bm{k} 
\end{equation}
and
\begin{equation}
 \label{Naeherung2}
\bm{J}_1=-\int\limits_0^T\!\mathrm{d}t\,t^2 \Gamma(\overline{\bm{r}})\bm{k} -\int\limits_T^{2T}\!\mathrm{d}t\, (2T-t)t \Gamma(\overline{\bm{r}})\bm{k} 
\end{equation}
derived in the Supplemental material of
Ref.~\cite{Compensation2} 
after appropriate resummation of the integrals.

In \Eq{TaylorDecomposition} corrections from the next order of the Taylor expansion are suppressed by $v_r T/\xi$ where $\xi$, the characteristic length scale on which the potential changes.
For the local variations in the gravitational profile from Ref.~\cite{VerticalProfile} the factor $v_r T/\xi$ might approach unity in future experiments involving large-momentum transfer techniques and therefore limits the validity of Eqs.~\eqref{Naeherung1} and \eqref{Naeherung2}.
Instead, using the midpoint theorem \cite{Borde3} without the approximation in \Eq{TaylorDecomposition}, the result still deviates from the exact expressions in \Eq{DefinitionJ} but only by a factor $(v_r T/\xi)^2$ which justifies its application in many cases but care has to be taken when employing LMT techniques or the potential changes on short length scales. 
This deviation results from the semiclassical approximation in the derivation of the  midpoint theorem  which limits its application to anharmonic potentials.

\subsection{Global gravity gradients}
So far we have discussed general anharmonic perturbations which might even change rapidly over the branch separation. 
In this paragraph we assume that the deviations from linear gravity are smooth enough to be accurately described over the extent of the whole interferometer by a \textit{global} gradient. Correspondingly,
\begin{equation}
\label{Gravitational potential}
    V=\frac{1}{2}m \bm{r}^\mathrm{T}\Gamma \bm{r}
\end{equation}
where the position-independent gradient tensor $\Gamma$ is chosen fully symmetric. In case all laser pulses are aligned, we define the vector $ \bm{A}=\oint\! \mathrm{d} t\,  \bm{r}_0$, whose modulus corresponds to the space-time area enclosed by the two branches of the unperturbed interferometer. 
With 
$\bm{a}=-\Gamma\bm{r}_0$ the expression 
\begin{equation}
\label{centroid}
    \bm{J}_0 t_c=\bm{J}_1
\end{equation}
defines the position of the geometric center $t_c$ of this area on the time axis and we can distinguish two different situations corresponding to the two classes of interferometer geometries displayed in \Fig{fig:Centeroid}: a) Suppose the interferometer exhibits a laser pulse at $t_1=t_c$ as for example in the MZ interferometer visualized in \Fig{fig:Centeroid} a). From \Eq{ks} together with \Eq{centroid} we find 
\begin{equation}
    \Delta \bm{k}_1=\frac{m}{\hbar} \Gamma \bm{A}\,,\;\;\;\text{and}\;\;\;\Delta \bm{k}_2=0
\end{equation}
which agrees with the result of Ref.~\cite{Compensation1} for an MZ interferometer with $\bm{A}=\bm{v}_r T$ where the modification of momentum transfer is distributed equally over both branches. Thus, in case of uniform gradients, compensation is particularly simple if the interferometer exhibits a laser pulse at the geometric center of the space-time area enclosed by the branches. b) In contrast if $t_c$ is located exactly in between two pulses at $t_1$ and $t_2$, that is $t_c=(t_1+t_2)/2=T_1+T_2/2$,  we find
\begin{equation}
    \Delta \bm{k}_1= \Delta \bm{k}_2=\frac{m}{2\hbar}\Gamma\bm{A}\,.
\end{equation}
This situation is for example found in a Ramsey-Bord\'{e} interferometer shown in \Fig{fig:Centeroid} b) for which we find $\bm{A}=\bm{v}_rT_1(T_1+T_2)$.

\section{Symmetric trajectories}
\label{Special form of gradients}
In this appendix we consider interferometer schemes symmetric around a pulse at time $t_s$, that is $\bm{r}_0(t_s+t)=\bm{r}_0(t_s-t)$ for both branches. Since a function $q(t)$ symmetric around time $t_s$ satisfies
\begin{equation}
    \int_0^{2t_s}\! \mathrm{d}t\, q(t)t=t_s\int_0^{2t_s}\! \mathrm{d}t\, q(t)\,,
\end{equation}
we conclude from \Eq{DefinitionJ} that 
\begin{equation}
\label{centerofArea}
    \bm{J}_0 t_s=\bm{J}_1\,.
\end{equation}
Consequently, with the help of \Eq{ks} the compensation scheme simplifies to 
\begin{equation}
\label{Simplified solution}
 \Delta \bm{k}_1=-\bm{J}_0\quad\quad\text{and}\quad\quad\Delta \bm{k}_2=0
\end{equation}
for e.g.~$t_1=t_s$, so that only the pulse at this time must be modified.
A geometry satisfying this symmetry requirement is realized for example in an MZ interferometer with initial velocity  $v_{0z}=gT-v_r/2$ of the atoms in direction of the subsequent momentum transfer.
This result remains a good approximation if the branches are only approximately symmetric \cite{Compensation2}.

\end{document}